\pdfoutput=1

\documentclass[12pt,a4paper]{article}

\usepackage{ifthen}
\newboolean{pdflatex}
\setboolean{pdflatex}{true}

\newboolean{articletitles}
\setboolean{articletitles}{true}

\newboolean{uprightparticles}
\setboolean{uprightparticles}{false}

\newboolean{inbibliography}
\setboolean{inbibliography}{false}

\usepackage{microtype}
\usepackage{lineno}  
\usepackage{xspace} 

\usepackage{graphicx}  
\usepackage{color}
\usepackage{colortbl}
\graphicspath{{./figs/}} 

\usepackage{amsmath} 
\usepackage{amssymb}
\usepackage{amsfonts}
\usepackage{upgreek} 

\newcommand*\patchAmsMathEnvironmentForLineno[1]{%
\expandafter\let\csname old#1\expandafter\endcsname\csname #1\endcsname
\expandafter\let\csname oldend#1\expandafter\endcsname\csname
end#1\endcsname
 \renewenvironment{#1}%
   {\linenomath\csname old#1\endcsname}%
   {\csname oldend#1\endcsname\endlinenomath}%
}
\newcommand*\patchBothAmsMathEnvironmentsForLineno[1]{%
  \patchAmsMathEnvironmentForLineno{#1}%
  \patchAmsMathEnvironmentForLineno{#1*}%
}
\AtBeginDocument{%
\patchBothAmsMathEnvironmentsForLineno{equation}%
\patchBothAmsMathEnvironmentsForLineno{align}%
\patchBothAmsMathEnvironmentsForLineno{flalign}%
\patchBothAmsMathEnvironmentsForLineno{alignat}%
\patchBothAmsMathEnvironmentsForLineno{gather}%
\patchBothAmsMathEnvironmentsForLineno{multline}%
}

\usepackage{hyperref}    
\usepackage[all]{hypcap} 

\def\lhcb {\mbox{LHCb}\xspace}

\def\belle  {\mbox{Belle}\xspace}

\def\aleph  {\mbox{ALEPH}\xspace}

\ifthenelse{\boolean{uprightparticles}}%
{

 \def\Ppi         {\ensuremath{\uppi}\xspace}

 \def\PDelta      {\ensuremath{\Delta}\xspace}                 
 \def\PXi      {\ensuremath{\Xi}\xspace}                 
 \def\PLambda      {\ensuremath{\Lambda}\xspace}                 
 \def\PSigma      {\ensuremath{\Sigma}\xspace}                 
 \def\POmega      {\ensuremath{\Omega}\xspace}                 
 \def\PUpsilon      {\ensuremath{\Upsilon}\xspace}                 
 
 \def\PB      {\ensuremath{\mathrm{B}}\xspace}                 
                  
 \def\PD      {\ensuremath{\mathrm{D}}\xspace}

 \def\PK      {\ensuremath{\mathrm{K}}\xspace}

 \def\Pb      {\ensuremath{\mathrm{b}}\xspace}                 
 \def\Pc      {\ensuremath{\mathrm{c}}\xspace}

 \def\Pi      {\ensuremath{\mathrm{i}}\xspace}

 \def\Pp      {\ensuremath{\mathrm{p}}\xspace}

 \def\Ps      {\ensuremath{\mathrm{s}}\xspace}

}
{

 \def\Ppi         {\ensuremath{\pi}\xspace}

 \mathchardef\PDelta="7101
 \mathchardef\PXi="7104
 \mathchardef\PLambda="7103
 \mathchardef\PSigma="7106
 \mathchardef\POmega="710A
 \mathchardef\PUpsilon="7107
                  
 \def\PB      {\ensuremath{B}\xspace}                 
                  
 \def\PD      {\ensuremath{D}\xspace}

 \def\PK      {\ensuremath{K}\xspace}

 \def\Pb      {\ensuremath{b}\xspace}                 
 \def\Pc      {\ensuremath{c}\xspace}

 \def\Pi      {\ensuremath{i}\xspace}

 \def\Pp      {\ensuremath{p}\xspace}

 \def\Ps      {\ensuremath{s}\xspace}

}

\def\squark    {\ensuremath{\Ps}\xspace}

\def\cquark    {\ensuremath{\Pc}\xspace}

\def\bquark    {\ensuremath{\Pb}\xspace}

\def\pion  {\ensuremath{\Ppi}\xspace}
\def\piz   {\ensuremath{\pion^0}\xspace}

\def\pip   {\ensuremath{\pion^+}\xspace}
\def\pim   {\ensuremath{\pion^-}\xspace}

\def\kaon  {\ensuremath{\PK}\xspace}
  \def\Kbar  {\kern 0.2em\overline{\kern -0.2em \PK}{}\xspace}

\def\Kzb   {\ensuremath{\Kbar^0}\xspace}
\def\Kp    {\ensuremath{\kaon^+}\xspace}
\def\Km    {\ensuremath{\kaon^-}\xspace}

  \def\Dbar    {\kern 0.2em\overline{\kern -0.2em \PD}{}\xspace}
\def\D       {\ensuremath{\PD}\xspace}

\def\Dz      {\ensuremath{\D^0}\xspace}

\def\Dstarp  {\ensuremath{\D^{*+}}\xspace}

\def\B       {\ensuremath{\PB}\xspace}
\def\Bbar    {\ensuremath{\kern 0.18em\overline{\kern -0.18em \PB}{}}\xspace}

\def\Bz      {\ensuremath{\B^0}\xspace}

\def\Bu      {\ensuremath{\B^+}\xspace}

\def\Bd      {\ensuremath{\B^0}\xspace}
\def\Bs      {\ensuremath{\B^0_\squark}\xspace}

  \def\Y#1S{\ensuremath{\PUpsilon{(#1S)}}\xspace}

\def\proton      {\ensuremath{\Pp}\xspace}
\def\antiproton  {\ensuremath{\overline \proton}\xspace}

\def\Lz {\ensuremath{\PLambda}\xspace}
\def\Lbar {\ensuremath{\kern 0.1em\overline{\kern -0.1em\PLambda}}\xspace}

\def\Lb      {\ensuremath{\Lz^0_\bquark}\xspace}

\def\Lc      {\ensuremath{\Lz^+_\cquark}\xspace}

\def\BF         {{\ensuremath{\cal B}\xspace}}

\newcommand{\decay}[2]{\ensuremath{#1\!\to #2}\xspace}         

\def\to                 {\ensuremath{\rightarrow}\xspace}

\def\AT#1     {\ensuremath{A_{\mathrm{T}}^{#1}}\xspace}           

\def\C#1      {\ensuremath{\mathcal{C}_{#1}}\xspace}                       
\def\Cp#1     {\ensuremath{\mathcal{C}_{#1}^{'}}\xspace}                    
\def\Ceff#1   {\ensuremath{\mathcal{C}_{#1}^{\mathrm{(eff)}}}\xspace}        
\def\Cpeff#1  {\ensuremath{\mathcal{C}_{#1}^{'\mathrm{(eff)}}}\xspace}       
\def\Ope#1    {\ensuremath{\mathcal{O}_{#1}}\xspace}                       
\def\Opep#1   {\ensuremath{\mathcal{O}_{#1}^{'}}\xspace}                    

\newcommand{\tev}{\ifthenelse{\boolean{inbibliography}}{\ensuremath{~T\kern -0.05em eV}\xspace}{\ensuremath{\mathrm{\,Te\kern -0.1em V}}\xspace}}
\newcommand{\gev}{\ensuremath{\mathrm{\,Ge\kern -0.1em V}}\xspace}
\newcommand{\mev}{\ensuremath{\mathrm{\,Me\kern -0.1em V}}\xspace}
\newcommand{\kev}{\ensuremath{\mathrm{\,ke\kern -0.1em V}}\xspace}
\newcommand{\ev}{\ensuremath{\mathrm{\,e\kern -0.1em V}}\xspace}
\newcommand{\gevc}{\ensuremath{{\mathrm{\,Ge\kern -0.1em V\!/}c}}\xspace}
\newcommand{\mevc}{\ensuremath{{\mathrm{\,Me\kern -0.1em V\!/}c}}\xspace}
\newcommand{\gevcc}{\ensuremath{{\mathrm{\,Ge\kern -0.1em V\!/}c^2}}\xspace}
\newcommand{\gevgevcccc}{\ensuremath{{\mathrm{\,Ge\kern -0.1em V^2\!/}c^4}}\xspace}
\newcommand{\mevcc}{\ensuremath{{\mathrm{\,Me\kern -0.1em V\!/}c^2}}\xspace}

\def\mum  {\ensuremath{\,\upmu\rm m}\xspace}

\def\invfb   {\ensuremath{\mbox{\,fb}^{-1}}\xspace}

\newcommand{\chisq}{\ensuremath{\chi^2}\xspace}

\newcommand{\chisqip}{\ensuremath{\chi^2_{\rm IP}}\xspace}

\def\gsim{{~\raise.15em\hbox{$>$}\kern-.85em
          \lower.35em\hbox{$\sim$}~}\xspace}
\def\lsim{{~\raise.15em\hbox{$<$}\kern-.85em
          \lower.35em\hbox{$\sim$}~}\xspace}

\def\ptot       {\mbox{$p$}\xspace}
\def\pt         {\mbox{$p_{\rm T}$}\xspace}

\def\evtgen     {\mbox{\textsc{EvtGen}}\xspace}

\def\geant      {\mbox{\textsc{Geant4}}\xspace}

\def\photos     {\mbox{\textsc{Photos}}\xspace}

\def\pythia     {\mbox{\textsc{Pythia}}\xspace}

\def\tell1  {TELL1\xspace}
\def\ukl1   {UKL1\xspace}

\newcommand{\eg}{\mbox{\itshape e.g.}\xspace}
\newcommand{\ie}{\mbox{\itshape i.e.}\xspace}

\usepackage{cite} 
\usepackage{mciteplus}

\usepackage{longtable}

\usepackage{subfig}

\newcommand{\BPPbar}{\texorpdfstring{\decay{\Bz_{(s)}}{\proton \antiproton}}{}}
\newcommand{\BdPPbar}{\texorpdfstring{\decay{\Bd}{\proton \antiproton}}{}}
\newcommand{\BsPPbar}{\texorpdfstring{\decay{\Bs}{\proton \antiproton}}{}}
\newcommand{\BdKPi}{\texorpdfstring{\decay{\Bd}{\Kp \pim}}{}}
\newcommand{\BdPiPi}{\texorpdfstring{\decay{\Bd}{\pip \pim}}{}}

\newcommand{\BsKK}{\texorpdfstring{\decay{\Bs}{\Kp \Km}}{}}

\newcommand{\BsPiK}{\texorpdfstring{\decay{\Bs}{\pip \Km}}{}}
\newcommand{\LbPPi}{\texorpdfstring{\decay{\PLambda_b^0}{\proton \pim}}{}}
\newcommand{\LPPi}{\texorpdfstring{\decay{\PLambda}{\proton \pim}}{}}

\newcommand{\BdPiPiPiz}{\texorpdfstring{\decay{\Bd}{\pip \pim \piz}}{}}

\newcommand{\BdKKPiz}{\texorpdfstring{\decay{\Bd}{\Kp \Km \piz}}{}}

\newcommand{\BuPLbarExcited}{\texorpdfstring{\decay{\Bu}{\proton \Lbar(1520)}}{}}

\def\PPbar {$\proton\antiproton$\xspace}
\def\KPi   {\ensuremath{\Kp\pim}\xspace}

\newcommand{\erfc}[1]{\ensuremath{{\rm erfc}\Big(#1\Big)}\xspace}

\def\splot{\mbox{\em sPlot}\xspace}

\begin{document}

\renewcommand{\thefootnote}{\fnsymbol{footnote}}
\setcounter{footnote}{1}

\begin{titlepage}
\pagenumbering{roman}

\vspace*{-1.5cm}
\centerline{\large EUROPEAN ORGANIZATION FOR NUCLEAR RESEARCH (CERN)}
\vspace*{1.5cm}
\hspace*{-0.5cm}
\begin{tabular*}{\linewidth}{lc@{\extracolsep{\fill}}r}
\ifthenelse{\boolean{pdflatex}}
{\vspace*{-2.7cm}\mbox{\!\!\!\includegraphics[width=.14\textwidth]{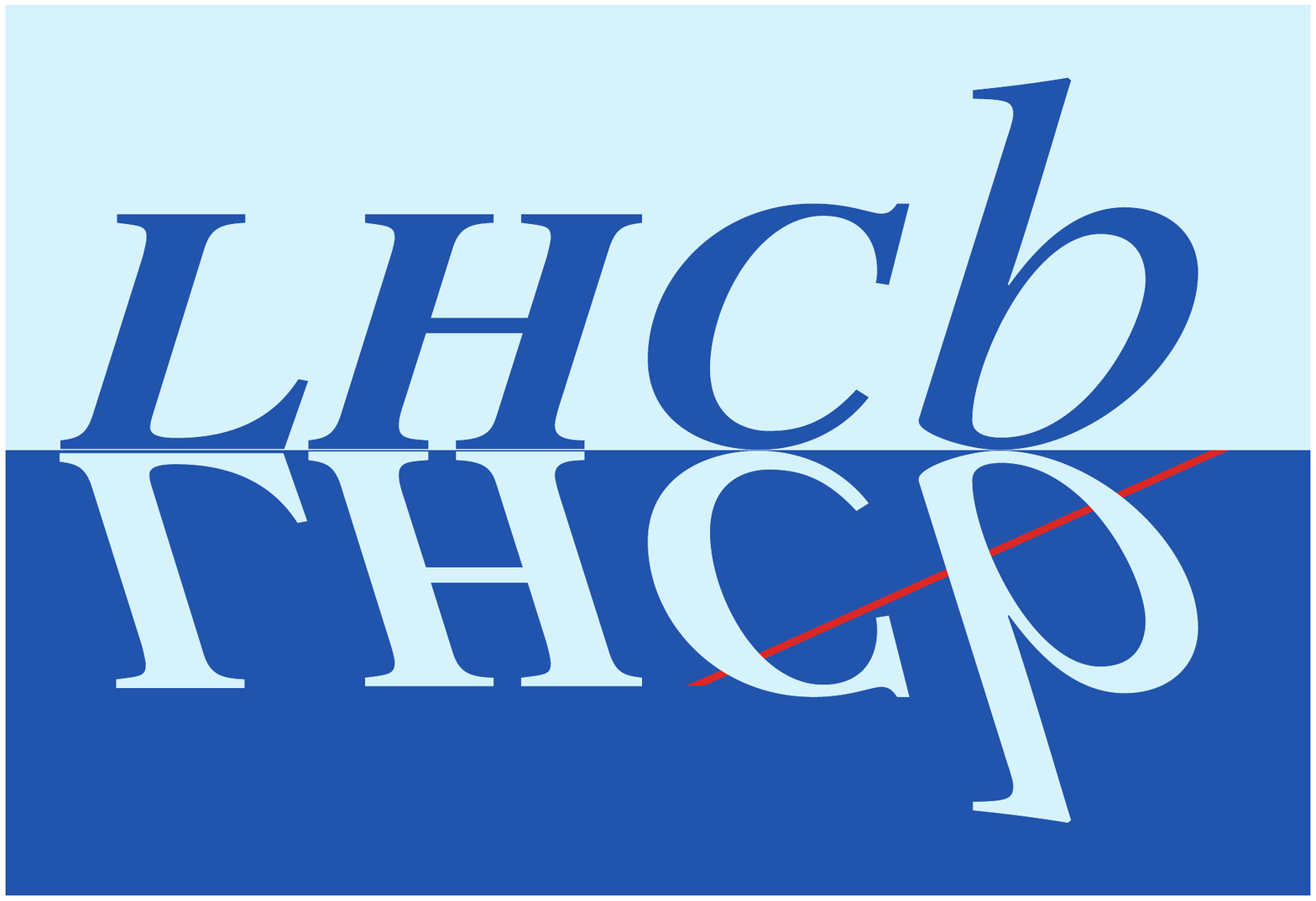}} & &}%
{\vspace*{-1.2cm}\mbox{\!\!\!\includegraphics[width=.12\textwidth]{lhcb-logo.eps}} & &}%
\\
 & & CERN-PH-EP-2013-138 \\
 & & LHCb-PAPER-2013-038 \\
 & & August 5, 2013 \\
 & & \\
\end{tabular*}

\vspace*{2.0cm}

{\bf\boldmath\huge
\begin{center}
First evidence for the two-body charmless baryonic decay \BdPPbar
\end{center}
}

\vspace*{2.0cm}

\begin{center}
The LHCb collaboration\footnote{Authors are listed on the following pages.}
\end{center}

\vspace{\fill}

\begin{abstract}
  \noindent
The results of a search for the rare two-body charmless baryonic decays
\BdPPbar and \BsPPbar are reported. The analysis uses a data sample,
corresponding to an integrated luminosity of 0.9~\invfb, of $pp$ collision
data collected by the LHCb experiment at a centre-of-mass energy of 7~\tev.
An excess of \BdPPbar candidates with respect to background expectations
is seen with a statistical significance of 3.3 standard deviations.
This is the first evidence for a two-body charmless baryonic \Bz decay.
No significant \BsPPbar signal is observed, leading to an improvement of
three orders of magnitude over previous bounds.
If the excess events are interpreted as signal,
the 68.3\% confidence level intervals on the branching fractions are
\begin{eqnarray}
\BF(\BdPPbar) & = & ( 1.47 \,^{+0.62}_{-0.51} \,^{+0.35}_{-0.14} ) \times 10^{-8} \,, \nonumber \\ \vspace*{0.3cm}
\BF(\BsPPbar) & = & ( 2.84 \,^{+2.03}_{-1.68} \,^{+0.85}_{-0.18} ) \times 10^{-8} \,, \nonumber
\end{eqnarray}
where the first uncertainty is statistical and the second is systematic.
\end{abstract}

\vspace*{2.0cm}

\begin{center}
  Submitted to JHEP
\end{center}

\vspace{\fill}

{\footnotesize 
\centerline{\copyright~CERN on behalf of the \lhcb collaboration, license \href{http://creativecommons.org/licenses/by/3.0/}{CC-BY-3.0}.}}
\vspace*{2mm}

\end{titlepage}

\newpage
\setcounter{page}{2}
\mbox{~}
\newpage

\centerline{\large\bf LHCb collaboration}
\begin{flushleft}
\small
R.~Aaij$^{40}$, 
B.~Adeva$^{36}$, 
M.~Adinolfi$^{45}$, 
C.~Adrover$^{6}$, 
A.~Affolder$^{51}$, 
Z.~Ajaltouni$^{5}$, 
J.~Albrecht$^{9}$, 
F.~Alessio$^{37}$, 
M.~Alexander$^{50}$, 
S.~Ali$^{40}$
% ...

{\footnotesize \it
% ...
$ ^{5}$Clermont Universit\'{e}, Universit\'{e} Blaise Pascal, CNRS/IN2P3, LPC, Clermont-Ferrand, France\\
$ ^{6}$CPPM, Aix-Marseille Universit\'{e}, CNRS/IN2P3, Marseille, France\\
% ...
$ ^{9}$Fakult\"{a}t Physik, Technische Universit\"{a}t Dortmund, Dortmund, Germany\\
% ...
$ ^{36}$Universidad de Santiago de Compostela, Santiago de Compostela, Spain\\
$ ^{37}$European Organization for Nuclear Research (CERN), Geneva, Switzerland\\
% ...
$ ^{40}$Nikhef National Institute for Subatomic Physics, Amsterdam, The Netherlands\\
% ...
$ ^{45}$H.H. Wills Physics Laboratory, University of Bristol, Bristol, United Kingdom\\
% ...
$ ^{50}$School of Physics and Astronomy, University of Glasgow, Glasgow, United Kingdom\\
$ ^{51}$Oliver Lodge Laboratory, University of Liverpool, Liverpool, United Kingdom\\
% ...
}
\end{flushleft}

\cleardoublepage

\renewcommand{\thefootnote}{\arabic{footnote}}
\setcounter{footnote}{0}

\pagestyle{plain}
\setcounter{page}{1}
\pagenumbering{arabic}

\linenumbers

\section{Introduction}
\label{sec:intro}
The observation of $B$ meson decays into two charmless mesons
has been reported in several decay modes~\cite{PDG2012}. Despite
various searches at $e^+ e^-$ colliders~\cite{Buskulic1996471,Coan:1998hy,Aubert:2004fy,Tsai:2007pp},
it is only recently that the LHCb collaboration reported the first observation
of a two-body charmless baryonic $B$ decay, the \BuPLbarExcited
mode~\cite{LHCb-PAPER-2013-031}.
This situation is in contrast with the observation of a multitude of
three-body charmless baryonic $B$ decays whose branching fractions are
known to be larger than those of the two-body modes; the former exhibit a
so-called threshold enhancement, with the baryon-antibaryon pair being
preferentially produced at low invariant mass, while the suppression of the
latter may be related to the same
effect~\cite{doi:10.1146/annurev.nucl.010909.083136}.

In this paper, a search for the \BdPPbar and \BsPPbar rare decay modes
at LHCb is presented. Both branching fractions are measured with
respect to that of the \BdKPi decay mode. The inclusion of charge-conjugate
processes is implied throughout this paper.

In the Standard Model (SM), the \BdPPbar mode decays via the $b \rightarrow u$
tree-level process whereas the penguin-dominated decay \BsPPbar is expected
to be further suppressed.
Theoretical predictions of the branching fractions for two-body
charmless baryonic \Bz decays within the SM vary depending on the
method of calculation used, \eg quantum chromodynamics sum rules, diquark model
and pole model. The predicted branching fractions are typically of order
$10^{-7}\!-\!10^{-6}$~\cite{Chernyak1990137,springerlink:10.1007/BF01548569,PhysRevD.43.1599,Jarfi1990513,PhysRevD.66.014020}.
No theoretical predictions have been published for the branching fraction
of two-body charmless baryonic decays of the \Bs meson.

The experimental 90$\%$ confidence level (CL) upper limit on the
\BdPPbar branching fraction, $\BF( \BdPPbar) < 1.1 \times 10^{-7}$,
is dominated by the latest search by the \belle experiment~\cite{Tsai:2007pp}
and has already ruled out most theoretical predictions. A single experimental
search exists for the corresponding \BsPPbar mode, performed by \aleph,
yielding the upper limit $\BF( \BsPPbar) < 5.9 \times 10^{-5}$
at 90\% CL~\cite{Buskulic1996471}.

\section{Detector and trigger}
\label{sec:detector}
The \lhcb detector~\cite{Alves:2008zz} is a single-arm forward
spectrometer covering the \mbox{pseudorapidity} range $2<\eta <5$,
designed for the study of particles containing \bquark or \cquark
quarks. The detector includes a high-precision tracking system
consisting of a silicon-strip vertex detector surrounding the $pp$
interaction region, a large-area silicon-strip detector located
upstream of a dipole magnet with a bending power of about
$4{\rm\,Tm}$, and three stations of silicon-strip detectors and straw
drift tubes placed downstream.
The combined tracking system provides momentum measurement with
relative uncertainty that varies from 0.4\% at 5\gevc to 0.6\% at 100\gevc,
and impact parameter (IP) resolution of 20\mum for
tracks with high transverse momentum (\pt). Charged hadrons are identified
using two ring-imaging Cherenkov detectors~\cite{LHCb-DP-2012-003}.
Photon, electron and hadron candidates are identified by a calorimeter system
consisting of scintillating-pad and preshower detectors, an electromagnetic
calorimeter and a hadronic calorimeter. Muons are identified by a
system composed of alternating layers of iron and multiwire
proportional chambers~\cite{LHCb-DP-2012-002}.
The trigger~\cite{LHCb-DP-2012-004} consists of a
hardware stage, based on information from the calorimeter and muon
systems, followed by a software stage, which applies a full event
reconstruction.

Events are triggered and subsequently selected in a similar way for both
\BPPbar signal modes and the normalisation channel \BdKPi.
The software trigger requires a two-track secondary vertex
with a large sum of track \pt and significant displacement from the primary
$pp$ interaction vertices~(PVs).
At least one track should have $\pt > 1.7\gevc$ and \chisqip with respect to
any primary interaction greater than 16, where \chisqip is defined as the
difference in \chisq from the fit of a given PV reconstructed with and without
the considered track. A multivariate algorithm~\cite{BBDT} is used for the
identification of secondary vertices consistent with the decay of a
\bquark hadron.

Simulated data samples are used for determining the relative detector and
selection efficiencies between the signal and the normalisation modes:
$pp$ collisions are generated using \pythia~6.4~\cite{Sjostrand:2006za} with a
specific \lhcb configuration~\cite{LHCb-PROC-2010-056};
decays of hadronic particles are described by \evtgen~\cite{Lange:2001uf},
in which final state radiation is generated using
\photos~\cite{Golonka:2005pn}; and the interaction of the generated particles
with the detector and its response are implemented using the \geant
toolkit~\cite{Allison:2006ve, *Agostinelli:2002hh} as described in
Ref.~\cite{LHCb-PROC-2011-006}.

\section{Candidate selection}
\label{sec:selection}
The selection requirements of both signal modes and the normalisation channel
exploit the characteristic topology of two-body decays and their
kinematics. All daughter tracks tend to have larger \pt compared to generic
tracks from light-quark background owing to the high $B$ mass,
therefore a minimum \pt requirement is imposed for all daughter candidates.
Furthermore, the two daughters form a secondary vertex (SV) displaced from the
PV due to the relatively long $B$ lifetime. The reconstructed $B$ momentum
vector points to its production vertex, the PV, which results in the $B$ meson
having a small IP with respect to the PV. This is in contrast with the
daughters, which tend to have a large IP with respect to the PV as they
originate from the SV, therefore a minimum \chisqip with respect to the PVs
is imposed on the daughters. The condition that the $B$ candidate comes from
the PV is further reinforced by requiring that the angle between the $B$
candidate momentum vector and the line joining the associated PV and the $B$
decay vertex ($B$ direction angle) is close to zero.

To avoid potential biases, \PPbar candidates with invariant mass within
$\pm 50 \mevcc$ ($\approx 3\sigma$) around the known \Bz and \Bs masses,
specifically the region $[ 5230, 5417 ] \mevcc$, are not examined until all
analysis choices are finalised. The final selection of \PPbar candidates
relies on a boosted decision tree~(BDT) algorithm~\cite{Breiman} as a
multivariate classifier to separate signal from background.
Additional preselection criteria are applied prior to the BDT training.

The BDT is trained with simulated signal samples and data from the sidebands
of the \PPbar mass distribution as background.
Of the $1.0 \invfb$ of data recorded in 2011,
10\% of the sample is exploited for the training of the \BPPbar selection,
and 90\% for the actual search.
The BDT training relies on an accurate description of the distributions of the
selection variables in simulated events. The agreement between
simulation and data is checked on the \BdKPi proxy decay with
distributions obtained from data using the \splot technique~\cite{Pivk:2004ty}.
No significant deviations are found, giving confidence that the inputs
to the BDT yield a nearly optimal selection. The variables used in the
BDT classifier are properties of the $B$ candidate and of the $B$ daughters,
\ie the proton and the antiproton.
The $B$ candidate variables are:
the vertex \chisq per number of degrees of freedom; the vertex \chisqip;
the direction angle; the distance in $z$ (the direction of the interacting
proton beams) between its decay vertex and the related PV;
and the \pt asymmetry within a cone around the $B$ direction defined by
$A_{\pt} = (\pt^{B} - \pt^{\text{cone}}) / (\pt^{B} + \pt^{\text{cone}})$,
with $\pt^{\text{cone}}$ being the \pt of the vector sum of the momenta of
all tracks measured within the cone radius $R=0.6$ around the $B$ direction,
except for the $B$-daughter particles.
The cone radius is defined in pseudorapidity and azimuthal angle
$( \eta, \phi)$ as $R = \sqrt{ (\Delta\eta)^{2} + (\Delta\phi)^{2} }$.
The BDT selection variables on the daughters are: their distance of closest
approach; the minimum of their \pt; the sum of their \pt; the minimum of their
\chisqip; the maximum of their \chisqip; and the minimum of their cone
multiplicities within the cone of radius $R = 0.6$ around them,
the daughter cone multiplicity being calculated as the number
of charged particles within the cone around each $B$ daughter.

The cone-related discriminators are motivated as isolation variables.
The cone multiplicity requirement ensures that the $B$ daughters are
reasonably isolated in space. The $A_{\pt}$
requirement further exploits the isolation of signal daughters in comparison to
random combinations of particles.

The figure of merit suggested in Ref.~\cite{punzi} is used to determine the
optimal selection point of the BDT classifier

\begin{equation}
  \text{FoM} = \frac{\epsilon^{\rm BDT}}
                    {a/2 + \sqrt{B_{\rm BDT}}} \,,
\label{eq:FoM}
\end{equation}
where $\epsilon^{\rm BDT}$ is the efficiency of the BDT selection on
the \BPPbar signal candidates, which is determined from simulation,
$B_{\rm BDT}$ is the expected number of background events within the
(initially excluded) signal region, estimated from the data sidebands, and the
term $a = 3$ quantifies the target level of significance in units of
standard deviation. With this optimisation the BDT classifier is found to
retain 44\% of the \BPPbar signals while reducing the combinatorial background
level by 99.6\%.

The kinematic selection of the \BdKPi decay is performed using
individual requirements on a set of variables similar to that used for the
BDT selection of the \BPPbar decays, except that
the cone variables are not used. This selection differs from the selection used
for signal modes and follows from the synergy with ongoing LHCb analyses on
two-body charmless $B$ decays, \eg Ref.~\cite{LHCb-PAPER-2012-013}.

The particle identification (PID) criteria applied in addition to the
\BPPbar BDT classifier are also optimised via Eq.~\ref{eq:FoM}. In this
instance, the signal efficiencies are determined from data control
samples owing to known discrepancies between data and simulation for the
PID variables. Proton PID efficiencies are tabulated in bins of
\ptot, \pt and the number of tracks in the event from data control
samples of \LPPi decays that are selected solely using kinematic
criteria. Pion and kaon efficiencies are likewise tabulated from data control
samples of $\Dstarp \to \Dz (\to \Km \pip)\,\pip$ decays.
The kinematic distributions of the simulated decay modes are then used to
determine an average PID efficiency.

Specific PID criteria are separately defined for the two signal modes and the
normalisation channel. The PID efficiencies are found to be
approximately 56\% for the \BPPbar signals and 42\% for \BdKPi decays.

The ratio of efficiencies of \BPPbar with respect to \BdKPi,
$\epsilon_{\BPPbar} / \epsilon_{\BdKPi}$,
including contributions from the detector acceptance, trigger,
selection and PID, is 0.60 (0.61). After all selection criteria are applied,
45 and 58009 candidates remain in the invariant mass ranges
$[ 5080, 5480 ] \mevcc$ and $[ 5000, 5800 ] \mevcc$ of the \PPbar and \KPi
spectra, respectively.

Possible sources of background to the \PPbar and \KPi spectra are investigated
using simulation samples. These include partially reconstructed backgrounds
with one or more particles from the decay of the $b$ hadron escaping detection,
and two-body $b$-hadron decays where one or both daughters are misidentified.

\section{Signal yield determination}
\label{sec:fits}
The signal and background candidates, in both the signal and normalisation
channels, are separated, after full selection, using unbinned maximum
likelihood fits to the invariant mass spectra.

The \KPi mass spectrum of the normalisation mode is described with a series
of probability density functions (PDFs) for the various components,
similarly to Ref.~\cite{LHCb-PAPER-2013-018}:
the \BdKPi signal, the \BsPiK signal, the \BsKK, \BdPiPi and the \LbPPi
misidentified backgrounds, partially reconstructed backgrounds, and
combinatorial background.
Any contamination from other decays is treated as a source of systematic
uncertainty.

Both signal distributions are modelled by the sum of two Crystal Ball (CB)
functions~\cite{Skwarnicki:1986xj} describing the high and low-mass
asymmetric tails. The peak values and the widths of the two CB components are
constrained to be the same.
All CB tail parameters and the relative normalisation of the two CB functions
are fixed to the values obtained from simulation whereas the signal peak value
and width are free to vary in the fit to the \KPi spectrum.
The \BsPiK signal width is constrained to the fitted \BdKPi width such that the
ratio of the widths is identical to that obtained in simulation.

The invariant mass distributions of the misidentified \BsKK, \BdPiPi and
\LbPPi backgrounds are determined from simulation and modelled with
non-parametric PDFs. The fractions of these misidentified backgrounds are
related to the fraction of the \BdKPi signal in the data via scaling factors
that take into account the relative branching fractions~\cite{PDG2012,HFAG},
$b$-hadron production fractions $f_q$~\cite{LHCb-PAPER-2011-018,Aaij:2013qqa},
and relevant misidentification rates.
The latter are determined from calibration data samples.

Partially reconstructed backgrounds represent decay modes that can
populate the spectrum when misreconstructed as signal with one or more
undetected final-state particles, possibly in conjunction with
misidentifications. The shape of this distribution is determined from
simulation, where each contributing mode is assigned a weight dependent on
its relative branching fraction, $f_q$ and selection efficiency.
The weighted sum of these partially-reconstructed backgrounds is shown
to be well modelled with the sum of two exponentially-modified Gaussian~(EMG)
functions
\begin{equation}
  \mbox{EMG}(x;\mu,\sigma,\lambda) = \frac{\lambda}{2}
  e^{\frac{\lambda}{2}(2x+\lambda\sigma^2-2\mu)} \cdot 
    \erfc{\frac{x+\lambda\sigma^2-\mu}{\sqrt{2}\sigma}} \,,
\end{equation}
where $\rm{erfc}(x) = 1 - \rm{erf}(x)$ is the complementary error function.
The signs of the variable $x$ and parameter $\mu$ are reversed
compared to the standard definition of an EMG function. The parameters
defining the shape of the two EMG functions and their relative weight
are determined from simulation. The component fraction of the
partially-reconstructed backgrounds is obtained from the fit to the data,
all other parameters being fixed from simulation.
The mass distribution of the combinatorial background is found to be well
described by a linear function whose gradient is determined
by the fit.

\begin{figure}[htbp]
  \centering
  \includegraphics[width=0.8\textwidth]{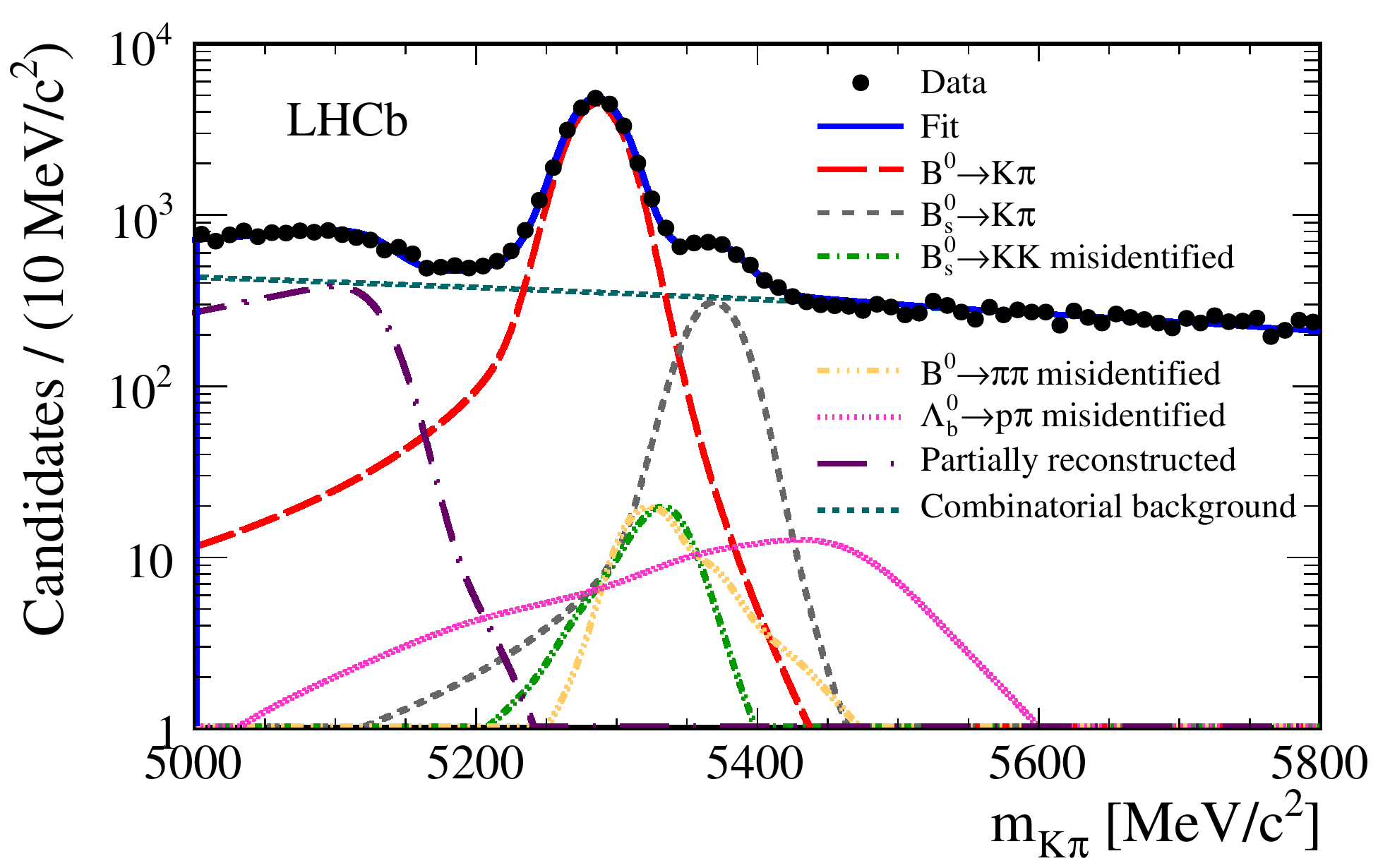}
  \includegraphics[width=0.8\textwidth]{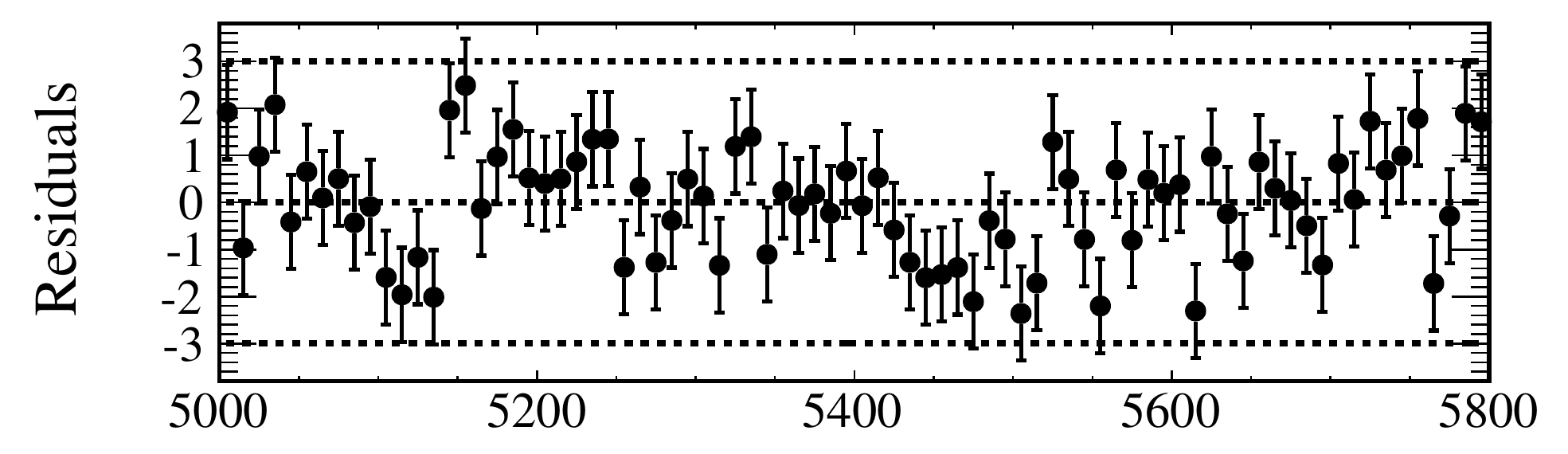}
  \caption{\small{Invariant mass distribution of \KPi candidates
      after full selection. The fit result (blue, solid) is superposed
      together with each fit model component as described in the
      legend. The normalised fit residual distribution is shown at
      the bottom.}}
	  \label{fig:fit_Kpi_spectrum}
\end{figure}

The fit to the \KPi spectrum, presented in Fig.~\ref{fig:fit_Kpi_spectrum},
determines seven parameters, and yields
$N(\BdKPi) = 24\,968 \pm 198$ signal events, 
where the uncertainty is statistical only.

\begin{figure}[htbp]
  \centering
  \includegraphics[scale=0.6]{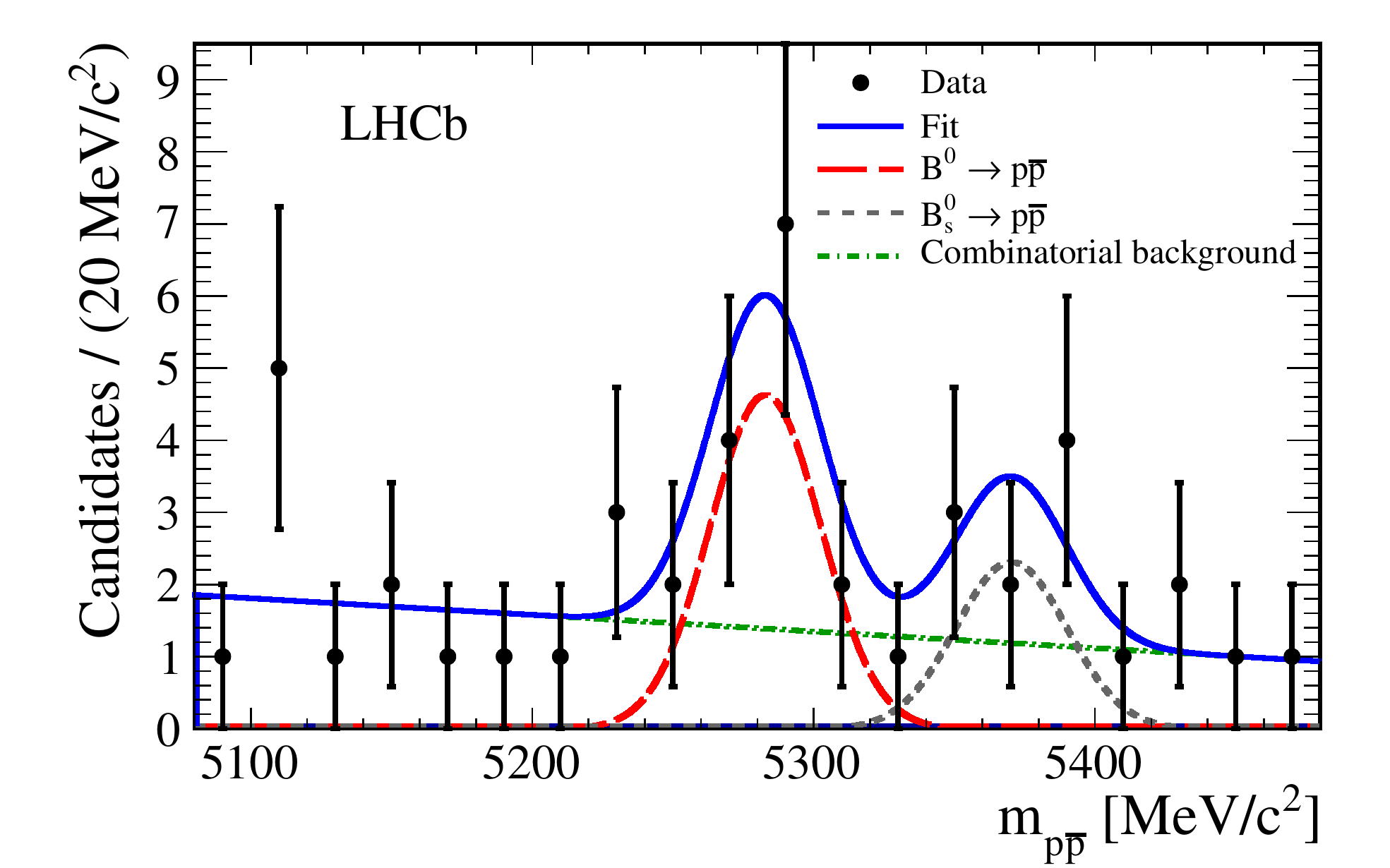}
  \caption{\small{Invariant mass distribution of \PPbar candidates
      after full selection. The fit result (blue, solid) is superposed
      with each fit model component: the \BdPPbar signal
      (red, dashed), the \BsPPbar signal (grey, dotted) and the combinatorial
      background (green, dot-dashed).}}
  \label{fig:fit_ppbar_spectrum}
\end{figure}

The \PPbar spectrum is described by PDFs for the three
components: the \BdPPbar and \BsPPbar signals, and the combinatorial
background. In particular, any contamination from partially
reconstructed backgrounds, with or without  misidentified
particles, is treated as a source of systematic uncertainty.

Potential sources of non-combinatorial background to the \PPbar
spectrum are two- and three-body decays of $b$ hadrons into protons, pions
and kaons,
and many-body $b$-baryon modes partially reconstructed,
with one or multiple misidentifications. It is verified from extensive
simulation studies that the ensemble of specific backgrounds do not peak
in the signal region but rather contribute to a smooth mass spectrum,
which can be accommodated by the dominant combinatorial background
contribution. The most relevant backgrounds are found to be
$\Lb \to \Lc (\to \proton \Kzb) \pim$,
$\Lb \to \Kzb \proton \pim$, \BdKKPiz and \BdPiPiPiz decays.
Calibration data samples are exploited to determine the PID efficiencies
of these decay modes, thereby confirming the suppression with respect to
the combinatorial background by typically one or two orders of magnitude.
Henceforth physics-specific backgrounds are neglected in the fit to the
\PPbar mass spectrum.

The \BPPbar signal mass shapes are verified in simulation to be well
described by a single Gaussian function. The widths of both Gaussian functions
are assumed to be the same for \BdPPbar and \BsPPbar; a systematic
uncertainty associated to this assumption is evaluated. They are determined
from simulation with a scaling factor to account for differences in the
resolution between data and simulation; the scaling factor is determined
from the \BdKPi data and simulation samples.
The mean of the \BsPPbar Gaussian function is constrained according to the
\Bs--\Bd mass difference~\cite{PDG2012}. The mass distribution of the
combinatorial background is described by a linear function.

The fit to the \PPbar mass spectrum is presented in
Fig.~\ref{fig:fit_ppbar_spectrum}.
The yields for the \BPPbar~signals in the full mass range are
$N(\BdPPbar) = 11.4 ^{+4.3}_{-4.1}$ and $N(\BsPPbar) = 5.7 ^{+3.5}_{-3.2}$,
where the uncertainties are statistical only.

\begin{figure}[tbp]
  \centering
  \includegraphics[width=0.48\textwidth]{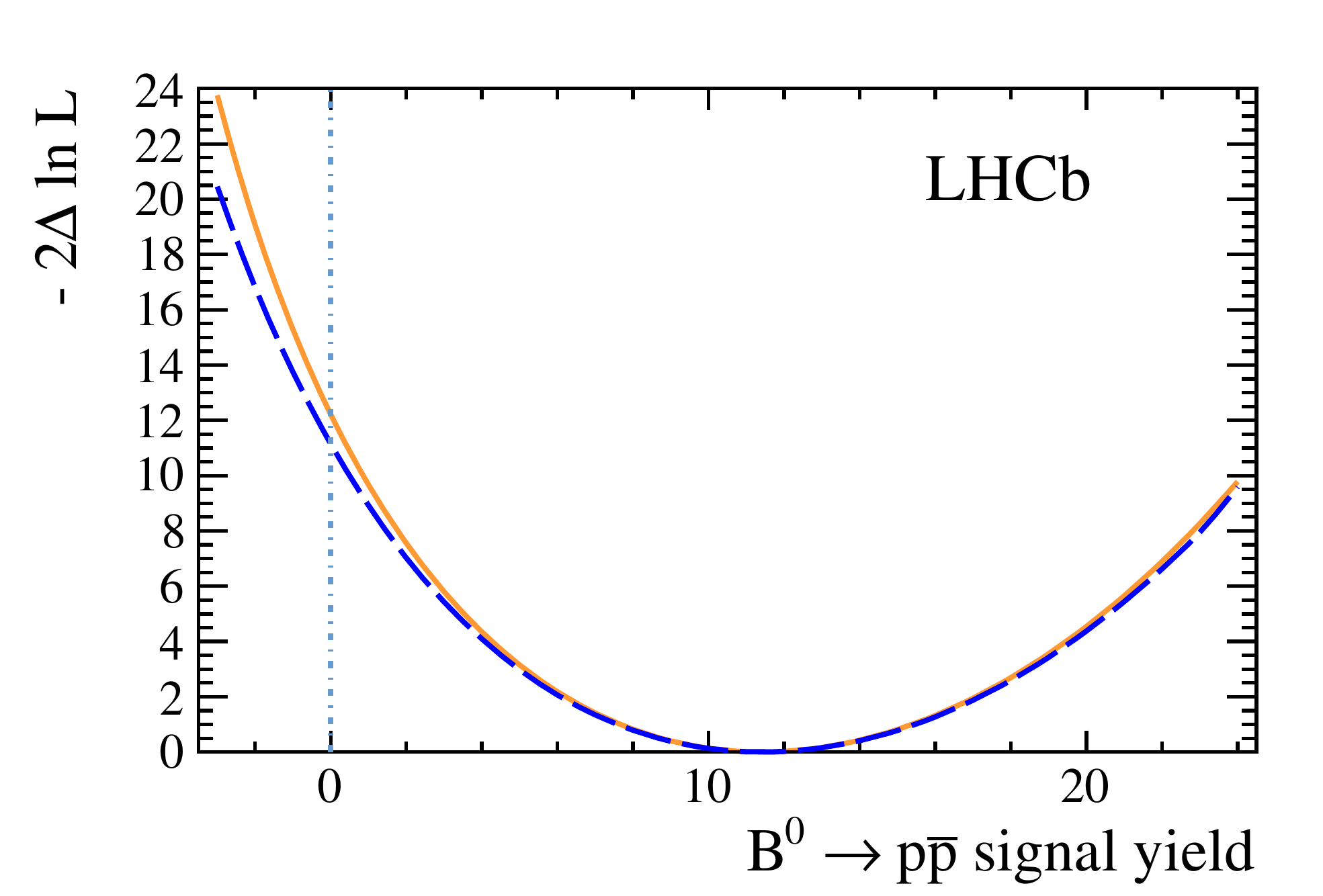}
  \includegraphics[width=0.48\textwidth]{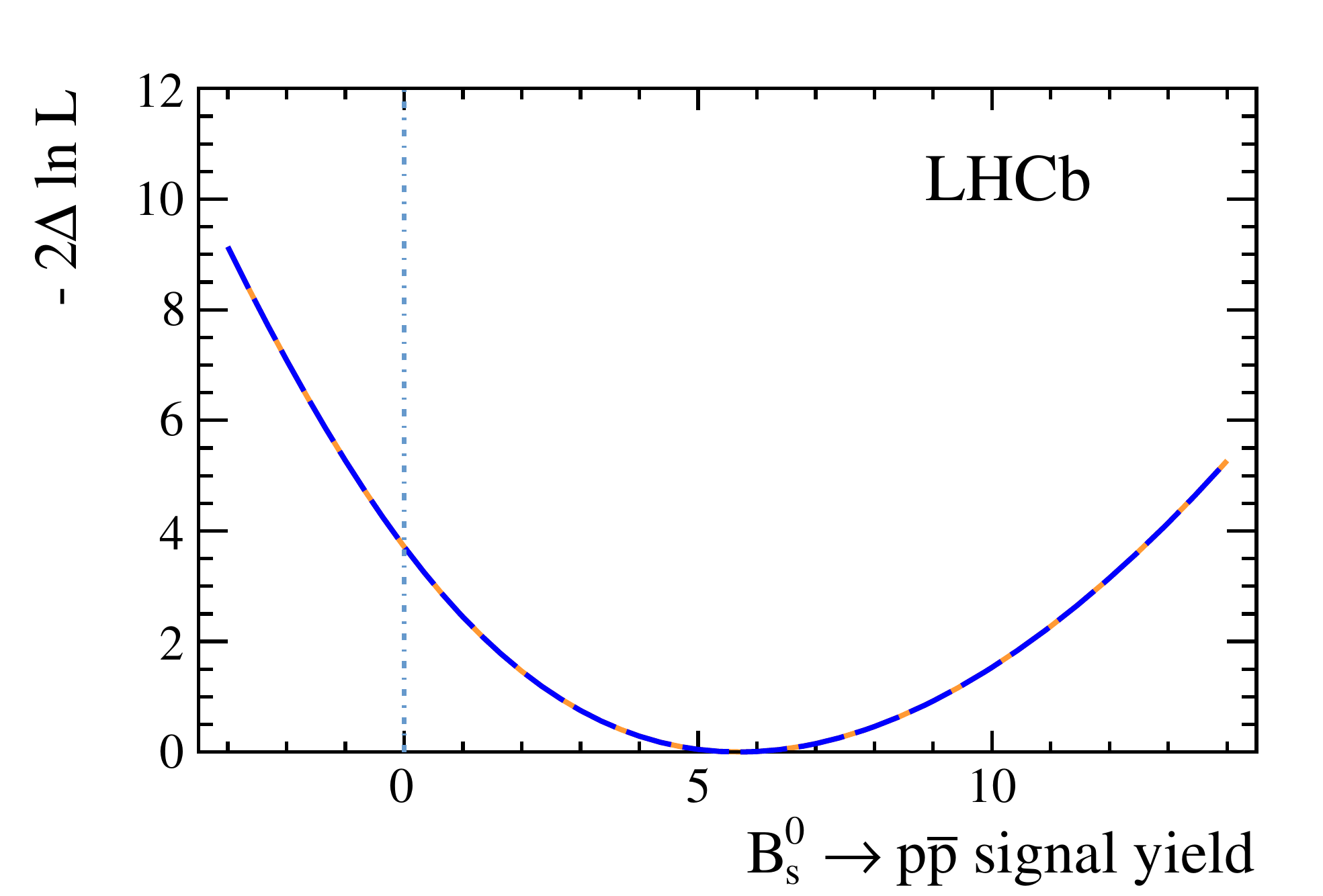}
  \caption{\small{Negative logarithm of the profile likelihoods
      as a function of (left) the \BdPPbar signal yield and
      (right) the \BsPPbar signal yield. The orange solid curves correspond to
      the statistical-only profiles whereas the blue dashed curves include
      systematic uncertainties.}}
  \label{fig:significances}
\end{figure}

The statistical significances of the \BPPbar signals are computed,
using Wilks' theorem~\cite{wilks1938large}, from the change in the
mass fit likelihood profiles when omitting the signal under scrutiny,
namely $\sqrt{2\ln(L_{\rm S+B}/L_{\rm B})}$, where $L_{\rm S+B}$ and
$L_{\rm B}$ are the likelihoods from the baseline fit and from the fit without
the signal component, respectively. The statistical significances are
$3.5\,\sigma$ and $1.9\,\sigma$ for the \BdPPbar and \BsPPbar decay modes,
respectively. Each statistical-only likelihood curve is convolved with
a Gaussian resolution function of width equal to the systematic uncertainty
(discussed below) on the signal yield.
The resulting likelihood profiles are presented in
Fig.~\ref{fig:significances}. The total signal significances are
$3.3\,\sigma$ and $1.9\,\sigma$ for the \BdPPbar and \BsPPbar modes,
respectively. We observe an excess of \BdPPbar candidates with
respect to background expectations; the \BsPPbar signal is not considered
to be statistically significant.

\section{Systematic uncertainties}
\label{sec:systematics}
The sources of systematic uncertainty are minimised by performing the
branching fraction measurement relative to a decay mode topologically
identical to the decays of interest. They are summarised in
Table~\ref{tab:syst_summary}.

\begin{table}[hbtp]
  \caption{\small{Relative systematic uncertainties contributing to the
           \BPPbar branching fractions.
           The total corresponds to the sum of all contributions added
           in quadrature.}}
  \begin{center}
    \vspace{\baselineskip}
    \begin{tabular}{ lccc}
      \hline
      Source                                  &  \multicolumn{3}{c}{Value (\%)} \\
                                              &  \BdPPbar & \BsPPbar & \BdKPi \\
      \hline
      \BdKPi branching fraction               &  --       & --       & 2.8 \\
      Trigger efficiency relative to \BdKPi   &  2.0      & 2.0      & --  \\
      Selection efficiency relative to \BdKPi &  8.0      & 8.0      & --  \\
      PID efficiency                          &  10.6     & 10.7     & 1.0 \\
      Yield from mass fit                     &  6.8      & 4.6      & 1.6 \\
      $f_s/f_d$                               &  --       & 7.8      & --  \\
      \hline
      Total                                   &  15.1     & 16.3     & 3.4 \\
      \hline
      \end{tabular}			      
  \end{center}
  \label{tab:syst_summary}
\end{table}

The branching fraction of the normalisation channel \BdKPi,
$\BF(\BdKPi) = ( 19.55 \pm 0.54 ) \times 10^{-6}$~\cite{HFAG}, is known
to a precision of 2.8\%, which is taken as a systematic uncertainty.
For the measurement of the \BsPPbar branching fraction, an
extra uncertainty arises from the 7.8\% uncertainty on
the ratio of fragmentation fractions
\mbox{$f_s / f_d = 0.256 \pm 0.020$}~\cite{Aaij:2013qqa}.

The trigger efficiencies are assessed from simulation for all decay modes.
The simulation describes well the ratio of efficiencies of
the relevant modes that comprise the same number of tracks in the
final state. Neglecting small \ptot and \pt differences between the
\BdPPbar and \BsPPbar modes, the ratios of $\BdKPi/\BPPbar$ trigger
efficiencies should be consistent within uncertainties. The difference of about
2\% observed in simulation is taken as systematic uncertainty.

The \BdKPi mode is used as a proxy for the assessment of the
systematic uncertainties related to the selection; \BdKPi signal
distributions are obtained from data, using the \splot technique, for
a variety of selection variables. From the level of agreement between
simulation and data, a systematic uncertainty of 8\% is derived for the
\BPPbar selection efficiencies relative to \BdKPi.

The PID efficiencies are determined from data control samples. The
associated systematic uncertainties are estimated by repeating the
procedure with simulated control samples, the uncertainties being equal to
the differences observed betweeen data and simulation, scaled by the PID
efficiencies estimated with the data control samples.
The systematic uncertainties on the PID efficiencies are found to be
10.6\%, 10.7\% and 1.0\% for the \BdPPbar, \BsPPbar and \BdKPi decay modes,
respectively. The large uncertainties on the proton PID efficiencies arise
from limited coverage of the proton control samples in the kinematic region
of interest for the signal.

Systematic uncertainties on the fit yields arise from the limited
knowledge or the choice of the mass fit models, and from the uncertainties
on the values of the parameters fixed in the fits.
They are investigated by studying a large number of simulated datasets,
with parameters varying within their estimated uncertainties.
Combining all sources of uncertainty in quadrature, the uncertainties on the
\BdPPbar, \BsPPbar and \BdKPi yields are 6.8\%, 4.6\% and 1.6\%, respectively.

\section{Results and conclusion}
\label{sec:results_conclusion}

The branching fractions are determined relative to the \BdKPi
normalisation channel according to
\begin{eqnarray}
\BF(\BPPbar) & = & \frac{N(\BPPbar)}{N(\BdKPi)}
                   \cdot
                   \frac{\epsilon_{\BdKPi}}{\epsilon_{\BPPbar}}
                   \cdot f_d / f_{d(s)}
                   \cdot \BF(\BdKPi) \nonumber \\
             & = & \alpha_{d(s)} \cdot N(\BPPbar) \,,
  \label{eq:yield_norm}
\end{eqnarray}
where $\alpha_{d(s)}$ are the single-event sensitivities equal to
$( 1.31 \pm 0.18 ) \times 10^{-9}$ and $( 5.04 \pm 0.81 ) \times 10^{-9}$
for the \BdPPbar and \BsPPbar decay modes, respectively;
their uncertainties amount to 14\% and 16\%, respectively.

The Feldman-Cousins (FC) frequentist method~\cite{PhysRevD.57.3873}
is chosen for the calculation of the branching fractions.
The determination of the 68.3\% and 90\% CL bands is performed with simulation
studies relating the measured signal yields to branching fractions,
and accounting for systematic uncertainties.
The 68.3\% and 90\% CL intervals are

$$
\begin{array}{rcrc}
\BF(\BdPPbar) = ( 1.47 \,^{+0.62}_{-0.51} \,^{+0.35}_{-0.14} ) \times 10^{-8} & \mbox{at} & 68.3\% & \mbox{CL} \,, \vspace*{0.2cm} \\
\BF(\BdPPbar) = ( 1.47 \,^{+1.09}_{-0.81} \,^{+0.69}_{-0.18} ) \times 10^{-8} & \mbox{at} &    90\% & \mbox{CL} \,, \vspace*{0.2cm} \\
\BF(\BsPPbar) = ( 2.84 \,^{+2.03}_{-1.68} \,^{+0.85}_{-0.18} ) \times 10^{-8} & \mbox{at} & 68.3\% & \mbox{CL} \,, \vspace*{0.2cm} \\
\BF(\BsPPbar) = ( 2.84 \,^{+3.57}_{-2.12} \,^{+2.00}_{-0.21} ) \times 10^{-8} & \mbox{at} &    90\% & \mbox{CL} \,, \vspace*{0.2cm} \\
\end{array}
$$

\noindent
where the first uncertainties are statistical and the second are systematic.

In summary, a search has been performed for the rare two-body charmless
baryonic decays \BdPPbar and \BsPPbar using a data sample, corresponding to
an integrated luminosity of 0.9 \invfb, of $pp$ collisions collected at a
centre-of-mass energy of 7 \tev by the LHCb experiment.
The results allow two-sided confidence limits to be placed on the branching
fractions of both \BdPPbar~and \BsPPbar for the first time.
We observe an excess of \BdPPbar candidates with respect to background
expectations with a statistical significance of $3.3\,\sigma$.
This is the first evidence for a two-body charmless baryonic \Bz decay.
No significant \BsPPbar signal is observed and the present result improves
the previous bound by three orders of magnitude.

The measured \BdPPbar branching fraction is incompatible with all published
theoretical predictions by one to two orders of magnitude and motivates
new and more precise theoretical calculations of two-body charmless baryonic
$B$ decays.
An improved experimental search for these decay modes at LHCb with the full
2011 and 2012 dataset will help to clarify the situation, in particular
for the \BsPPbar mode.

\section*{Acknowledgements}
\noindent We express our gratitude to our colleagues in the CERN
accelerator departments for the excellent performance of the LHC. We
thank the technical and administrative staff at the LHCb
institutes. We acknowledge support from CERN and from the national
agencies: CAPES, CNPq, FAPERJ and FINEP (Brazil); NSFC (China);
CNRS/IN2P3 and Region Auvergne (France); BMBF, DFG, HGF and MPG
(Germany); SFI (Ireland); INFN (Italy); FOM and NWO (The Netherlands);
SCSR (Poland); MEN/IFA (Romania); MinES, Rosatom, RFBR and NRC
``Kurchatov Institute'' (Russia); MinECo, XuntaGal and GENCAT (Spain);
SNSF and SER (Switzerland); NAS Ukraine (Ukraine); STFC (United
Kingdom); NSF (USA). We also acknowledge the support received from the
ERC under FP7. The Tier1 computing centres are supported by IN2P3
(France), KIT and BMBF (Germany), INFN (Italy), NWO and SURF (The
Netherlands), PIC (Spain), GridPP (United Kingdom). We are thankful
for the computing resources put at our disposal by Yandex LLC
(Russia), as well as to the communities behind the multiple open
source software packages that we depend on.

\addcontentsline{toc}{section}{References}
\setboolean{inbibliography}{true}
\bibliographystyle{LHCb}
\bibliography{main,LHCb-PAPER,LHCb-CONF,LHCb-DP}

\end{document}